# A Practical Assessment of the Power Grid Inertia Constant Using PMUs

Mohammadali Hayerikhiyavi, Aleksandar Dimitrovski, *Senior Member, IEEE*

***Abstract*****-- Installation of phasor measurement units (PMUs) in a number of substations in the power grid can help assess a set of its values and parameters, in particular those related to the dynamics when disturbances occur in the system. Inertia constant of the power grid is one of the system stability related parameters that is essential for planning of the system spinning reserve. Estimates of the grid inertia constant require precise information of the system frequency at the time of disturbance. In this paper, an improved method for such estimate is presented, which is based on the derived data of frequency variations obtained from PMUs. In addition, a way to obtain the appropriate interval for the lowest error estimate is discussed. The method has been applied to assess the inertia constant of the Iranian power grid, by obtaining the rate of change of frequency after a disturbance in the defined interval, and the rate of specified power imbalance between electrical power input and output.**

## I. Introduction

One of the most important applications of Wide Area Monitoring System (WAMS) is online monitoring of system parameters in order to estimate the system state. Applications are diverse: from estimates of generated power, to backup protection of transmission lines, to system stability margins [1-3]. The main feature of WAMS are synchronized measurements, obtained from PMUs installed in some of the substations [4]. Some applications of WAMS in power systems introduced recently include modeling of distribution network loads, and estimation of zero sequence impedances of power lines used in backup protection for ground faults [5]. Also, a new method based on phasor measurements using discrete wavelet transform (DWT) has been introduced to determine the unstable low-frequency oscillation of the power system [6]. State estimation using data from PMUs is discussed in [7,8].

The importance of system dynamics and the ability to predict system frequency response following a disturbance requires knowledge of the inertia constant of the system. This parameter plays an essential role in maintaining the system stability [9,10]. In this context, swing equation has been applied to determine the required reserve capacity [11, 12]. Some proposed techniques require a large amount of data to calculate the inertia constant, turning it into a complicated and complex problem [13- 15].

This paper proposes a simpler method to estimate the inertia constant of either an individual generator or an entire multi-machine system based on the Swing Equation. The only difference between the two cases is that for an individual generator, the frequency and power measurements used are taken from the terminals of the corresponding generator. For the multi-machine case, the frequency of the center of inertia and the net system power imbalance are used. During a fault, frequency graphs from various locations and substations are modeled by a first-order curve in an appropriate interval, using PMU data. The gradient of that line defines the rate of change of frequency (ROCOF), which is used to obtain the inertia constant of the entire power system. The rationale for selection of a proper interval is explained in the sequel.

## II. Definition of Inertia

Power system inertia is related to the stored energy in the rotating masses connected to the power system. It can also be defined as the period of time in which the stored energy in the rotating masses could be used to supply the total rated power. This energy is one of the basic parameters of the power system [8, 9]. For an individual generator, it is defined by the moment of inertia of the rotating mass, $J$, and the rotational speed. It is well known from basic physics that the moment of inertia of a rotating mass can be calculated from the ratio of the stored kinetic energy and its rated speed squared. Because the rotational speed of the rotating mass of a generator continuously changes around the nominal speed, $\omega_r$, it is assumed that it always rotates with nominal (synchronous) speed. This assumption allows inertia to be redefined with the inertia constant, $H$, given by (1):

$$H = \frac{(1/2) J \omega_r^2}{S_r} \quad (1)$$

Equation (1) indicates that the inertia constant is a parameter of the generator, normalized by its rated power. Therefore, to achieve the equivalent inertia constant of several different generators in the power network, it is necessary to recalculate the corresponding inertia constant of each generator based on the system's base power, $S_b$. This way, when the inertia constant of each generator is normalized by the same rated power, the total inertia constant is obtained as a sum of each individual constants.

## III. Swing Equation

The Swing Equation defines the relation between the difference of the mechanical, $P_{mi}$, and the electrical power, $P_{ei}$, and the ROCOF, ${df_i}/_{dt}$ in ${\text{Hz}}/_{\text{sec}}$, for generator $i$ with an inertia constant of $H_i$ (sec), for the time immediately after a disturbance has occurred. In its basic form, it gives(16,17):

$$\frac{2H}{f_n}.\frac{df_i}{dt} = P_{mi} - P_{ei} = \Delta P_i \quad , \quad i = 1,2,\dots,N \qquad (2)$$

Equation (2) does not take into account any controller action (governor or AVR), load response to power and frequency changes, and spinning reserves.

### A. Application for an Individual Generator

Equation (2) can be used to directly estimate the inertia constant of a generator when reliable and accurate measurements of frequency are available, especially for the first derivative of frequency (ROCOF) and the real power unbalance. The required ROCOF in (2) is obtained by using discrete frequency data (from PMUs). The difference between the two adjacent frequency samples in Hertz (Hz) is divided by the time step between when the two samples to approximate ROCOF. The power imbalance of the generator can be directly calculated from the difference between the electrical and the mechanical power (during the disturbance, which can also be obtained using PMU data), converted in p.u. With these two values, the inertia constant is obtained directly from (2).

### B. Application for a Multi-Machine System

In this case, the goal is to obtain system equivalent frequency in order to assess the total inertia constant of the system. During an event with a very high imbalance, local frequency could be different for each or most of the generators.

In this paper, the system frequency is calculated based on the concept of the frequency of the center of inertia (COI). This frequency$(f_c)$ is affected by all generators' inertia and is defined as follows:

$$f_c = \frac{\sum_{i=1}^{N} H_i f_i}{\sum_{i=1}^{N} H_i} = \frac{1}{H_T}\sum_{i=1}^{n} H_i f_i \qquad (3)$$

where $H_T$ is the total inertia of the system. By combining (3) and (2), for a system with $N$ generators, we obtain (4):

$$\frac{2H_T}{f_n}.\frac{df_e}{dt} = \Delta P = \sum_{i=1}^{N} \Delta P_i \qquad (4)$$

In (4), $\Delta P$ represents the difference between the total electrical and mechanical powers for all generators (net amount of power imbalance in system). Therefore, the total system inertia can be found from (5):

$$H_T = \frac{f_n.\sum_{i=1}^{N} \Delta P_i}{2\,\frac{df_c}{dt}} \qquad (5)$$

The average frequency throughout the system is the approximation of the center of inertia frequency.

## IV. PMU (Phasor measurement unit)

One of the most important functions and requirements of a modern energy management system is state estimation based on real time measurements. Power system state is defined based on a set of positive-sequence voltage component values found simultaneously from bus measurements around the system. One of the promising solutions for power network real time monitoring are PMUs [1,10], which collect data with high accuracy and sampling by using the GPS system. They provide accurate information of three-phase voltages and lines currents in substations, transformers, and loads. Based on this information, positive-sequence components of voltages and currents are calculated accurately at the same time on the scale of microseconds, from which the corresponding phase angles can be derived.

In the proposed method, data derived from PMUs in the phasor data concentrator (PDC) are used to calculate the inertia constant of the system, which is evaluated after a disruption and obtained based on ROCOF.

### A. PMU System

Figure1 shows a PMU system model with its main components [11]. The corresponding measured values from current and voltage transformers are processed through an anti-aliasing filter where they are sampled with an adequate frequency. Anti-aliasing filter ensures that the sampling frequency is more than 2 times the nominal system frequency or it will not properly convert the analog signal to digital. [18,19]

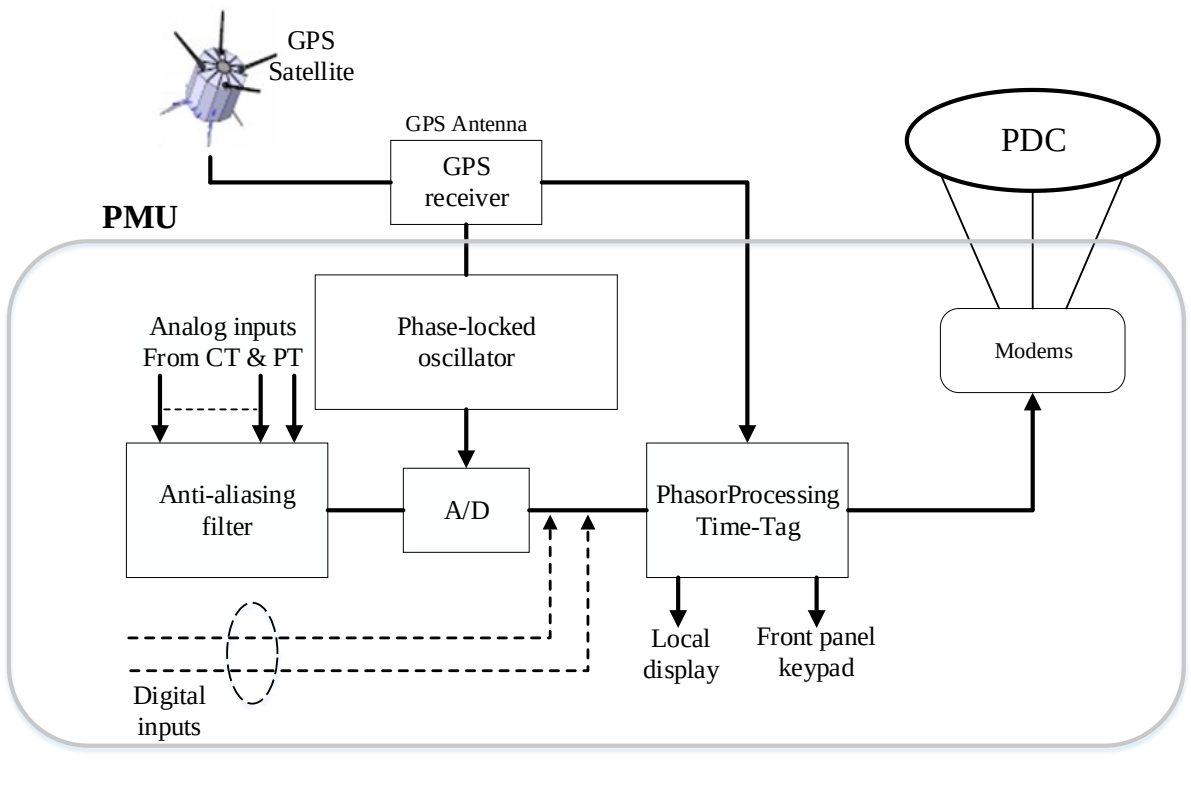

Fig. 1. PMU system model

## V. Interval Selection

Using the derived change of system frequency, the inertia constant can be calculated from (5). If these changes were obtained at the time of event occurrence, the measured inertia would not be accurate and reliable. PMUs have larger errors at the time of inception and accurate parameter estimates cannot be obtained. For example, in case of a loss of a power plant considered here, the remaining generators have to increase their power angle $\delta$ to supply the load, which can be approximately described using (6):

$$P = \frac{E.V}{X} sin\delta \quad (6)$$

where:
$E$: Voltage magnitude at the sending point
$V$: Voltage magnitude at the receiving point
$X$: Reactance between the sending and receiving point
$\delta$: Difference in angle between sending and receiving point

After the loss of generation, power angle immediately increases to supply the loads. Consequently, the frequency also immediately increases while, afterwards, the frequency of the system decreases since the loss of a part of generation is replenished from the kinetic energy of the remaining generators. The immediate increase in frequency at the time of generation loss can be considered as mechanical response of the system (in reaction to this event). Furthermore, rapid frequency changes occur immediately after the time of disturbance because of either stator transients ($\lambda'_d, \lambda'_q$) or back swing effects. If this period is used in the equation for estimating the constant of inertia, the result will have a large error. Our previous studies have shown that this error can be approximately 60%, and that the time from 0s (the moment of event inception) to about 1s should not be used for calculation of the inertia constant.

On the other hand, after the disturbance, automatic governor control systems get involved and try to restore the frequency back to the nominal value. These systems include mechanical elements with relatively high time constants. This will also cause another error in the calculation of the inertia constant of the system. In addition, for some loads, consumption decreases with frequency decline. This is the reason why in the case of a loss of generation frequency will not continue to decrease indefinitely (regardless of the governor and other control systems). Therefore, from approximately 4 seconds onward, the frequency data should not be taken into account.

In summary, in order to obtain the gradient of the frequency-time curve, the time from 0 to 1s, as well as from 4s onward after the event, should be discarded.

## VI. CASE STUDIES

### A. IEEE 39-bus

The proposed method is first applied on the IEEE 39-bus test system modelled in Digsilent software to demonstrate its validity and effectiveness. The one-line diagram of the test system is shown in Figure 2, and the generation inertia constants are given in Table I. By removing a certain generator from the system, while knowing the capacity of each generator, the inertia constant is calculated based on ROCOF. The calculation has been done twice in this system using simulated PMU data, once in the presence of the governors, and the second time without them.

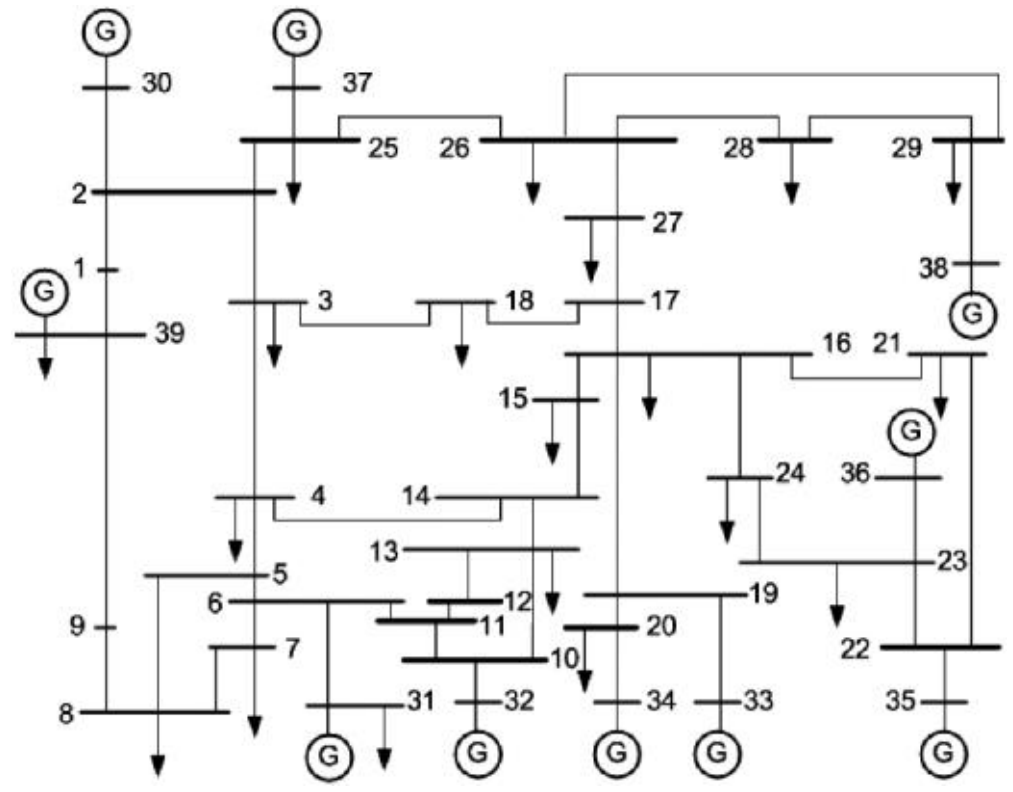

Fig. 2. IEEE 39 bus test system

Table I
Generation inertia constants for the IEEE 39-bus test system

| Generator ($G_i$) | Inertia Constant ($H_i$) (sec) | Actual Power(MW) |
|---|---|---|
| $G_{30}$ | 4.2 | 270 |
| $G_{31}$ | 4.329 | 585 |
| $G_{32}$ | 4.475 | 450 |
| $G_{33}$ | 3.575 | 632 |
| $G_{34}$ | 4.433 | 608 |
| $G_{35}$ | 4.35 | 1000 |
| $G_{36}$ | 3.771 | 560 |
| $G_{37}$ | 3.471 | 160 |
| $G_{38}$ | 3.45 | 245 |
| $G_{39}$ | 5 | 650 |

The system frequency plot is shown in Figure 3. It is the average frequency of all generators after the removal of the generator connected at bus 37 at simulation time 1s. Realistic governor control models have been implemented in the system for this case. However, for the sake of simplicity, AGC function was not implemented and the frequency at the end will not be restored to the rated value as it would be in an actual power system.

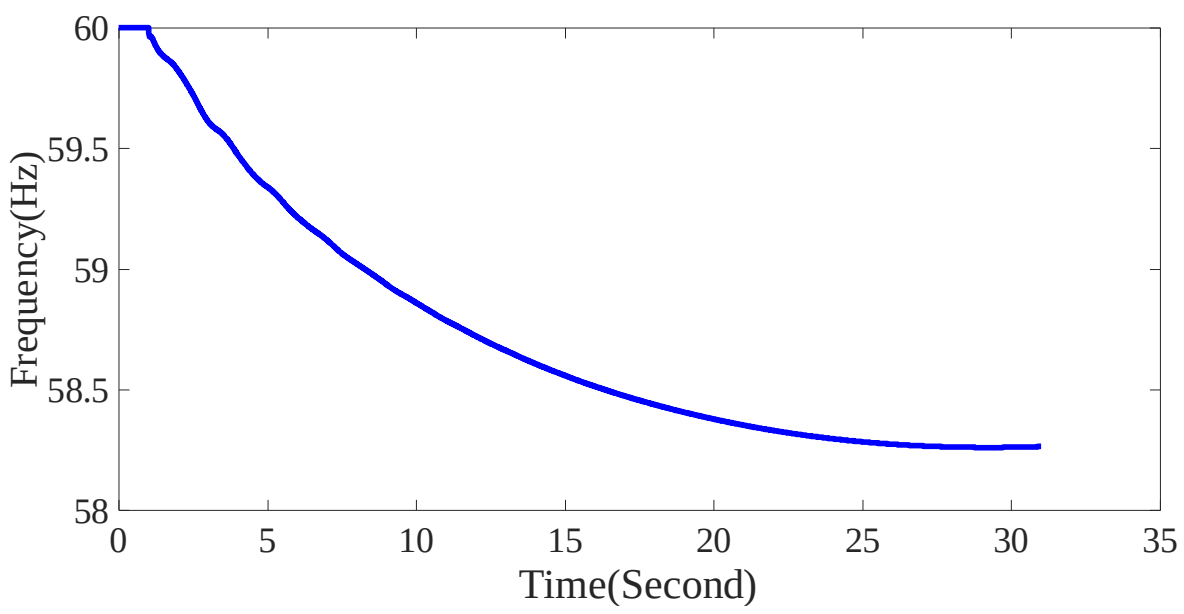

Fig. 3. Frequency after the removal of generator at bus 37 in presence of governors

A linear function, shown in Figure 4, is fit for the period from 2s to 5s after the event according to the discussion in Section V. Its slope in the interval of interest is 0.191, which corresponds to the ROCOF. Based on the proposed method, using the ROCOF and the generator powers in (5), the total inertia constant for this system is calculated at 4.38 s:

$$H = \frac{60 \times \frac{160}{5160 \div 0.9}}{2 \times 0.191} = 4.38$$

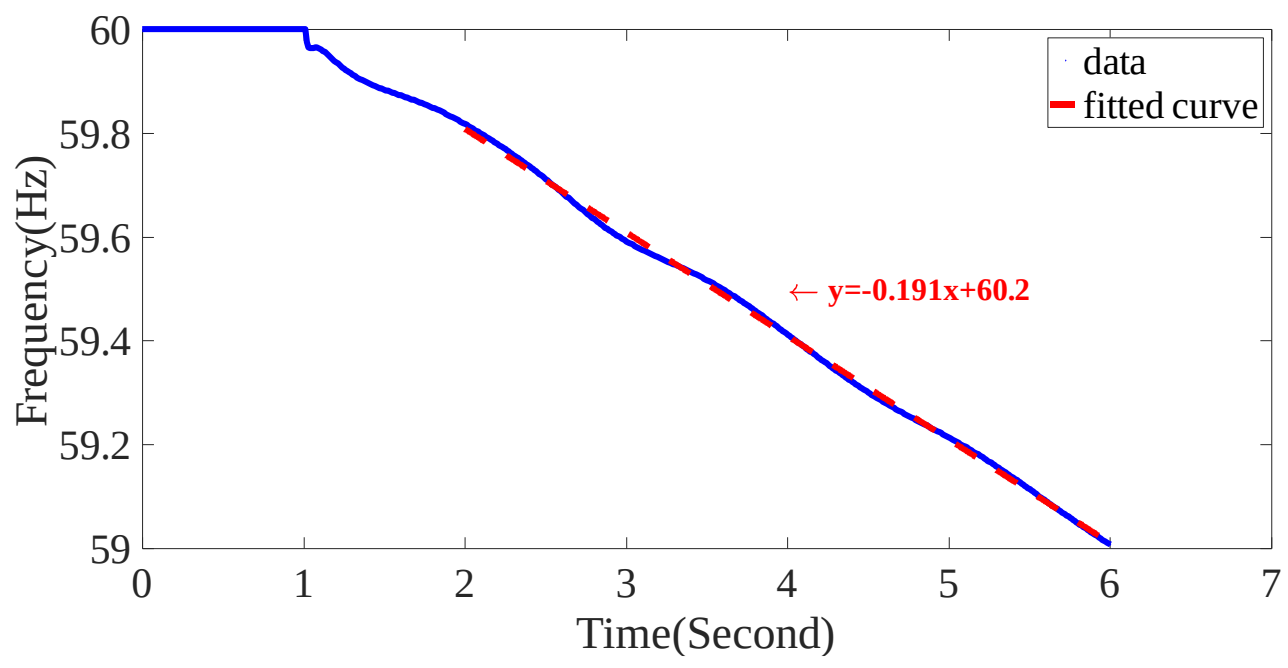

Fig. 4. Frequency curve in the interval from 2 to 5 s

In (5), the real power imbalance due to the event, $\Delta P$, must be in p.u in order to calculate the system inertia constant $H$ in seconds. Moreover, it must be noted that the power factor is 0.9. In general, it can be stated that the process of change of frequency initially depends on the $H$ of the system.

In order to demonstrate the suitability of the chosen interval for ROCOF calculation, the same case is repeated without governors. Figure 5 shows the plot of the frequency in this case. The slope of the curve from the onset of the event at 1s onward is 0.191.

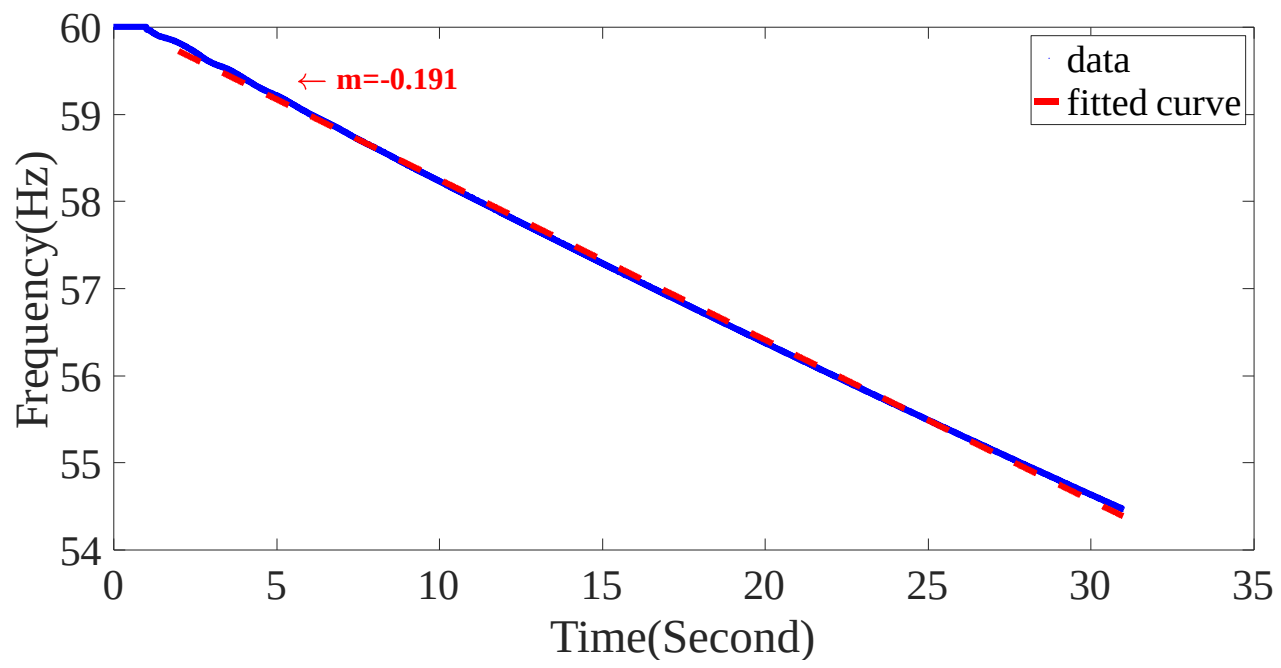

Fig. 5. Frequency after the removal of generator at bus 37 without governors

The permanent decrease of frequency is due to the absence of governors whose purpose is to arrest the frequency change (decline in this case).

The actual total system inertia constant $H_{Total}$, knowing the inertia constants of all the generators, can be calculated from:

$$H_{Total} = \frac{\sum H.MVA}{\sum MVA} \qquad (7)$$

Both sums in the above equation are over all the generators in the system including spinning reserves, except the part of the generation that has been removed after the incident. The actual value for the total system inertia of the IEEE 39-bus system based on the values shown in Table I and (7) is 4.3s.

## B. *Iranian power system*

The proposed method is applied again using data from a real event of a loss of a power plant in the Iranian power system. Figure 5 shows the change of frequency measured by PMUs in five substations. The samples are taken from the moment of the occurrence of an event related to a loss of part of the generation for 12 seconds after the incident. The samples were taken at 40 ms intervals. The horizontal axis shows the number of samples with 40 ms time step.

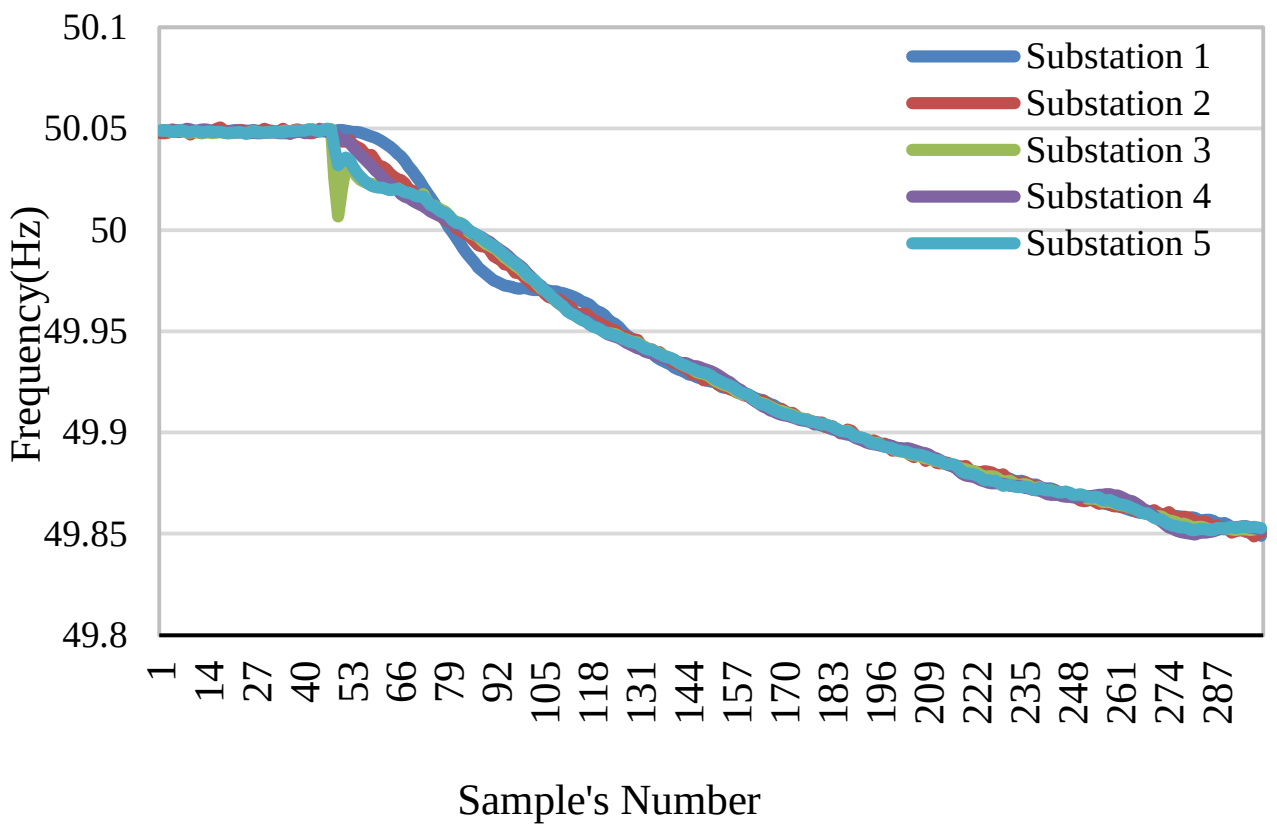

Fig. 6. Frequency data derived from PMUs installed in 5 substations after a disturbance

A zoomed-in plot of the frequency-time curve during the first second from the event inception is shown in Figure 7. As explained before, the time from 0s to 1s, as well as from 4s onward after the event, should be discarded.

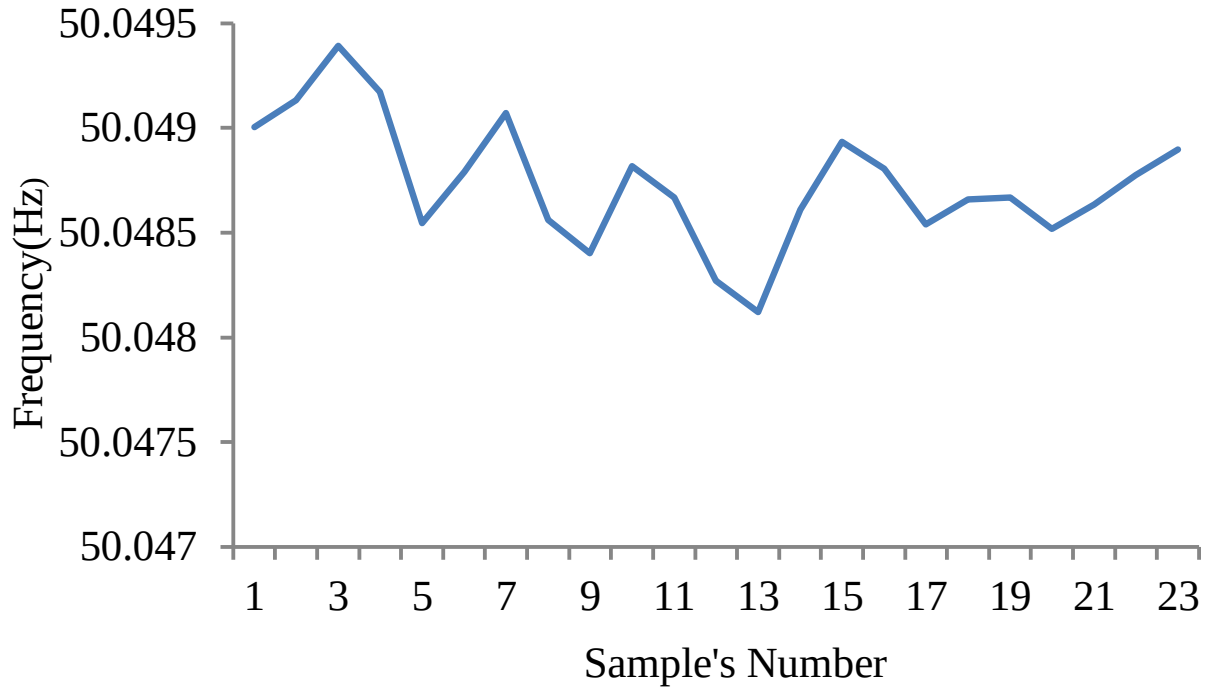

Fig. 7. Frequency curve in the interval from 0 to 1s

The frequency shown in Figure 7 is the average of the measurement from the five substations where the PMUs are located. It can be seen from the figure that, immediately after the incident, the frequency of the system increases and becomes higher than its initial value (before the event). The frequency decline starts from approximately 1s onwards.

As shown in Figure 8, a linear regression curve is fit for the period from 1s to 4s after the event. As stated before, the slope of the line corresponds to the ROCOF. According to the fitted line in the figure, the ROCOF in the interval of interest is 0.02.

The power related data for this event in the system are given in Table II.

Table II
PMU data

| Loss (MW) | Load before the disturbance(MW) |
|---|---|
| 200 | 20000 |

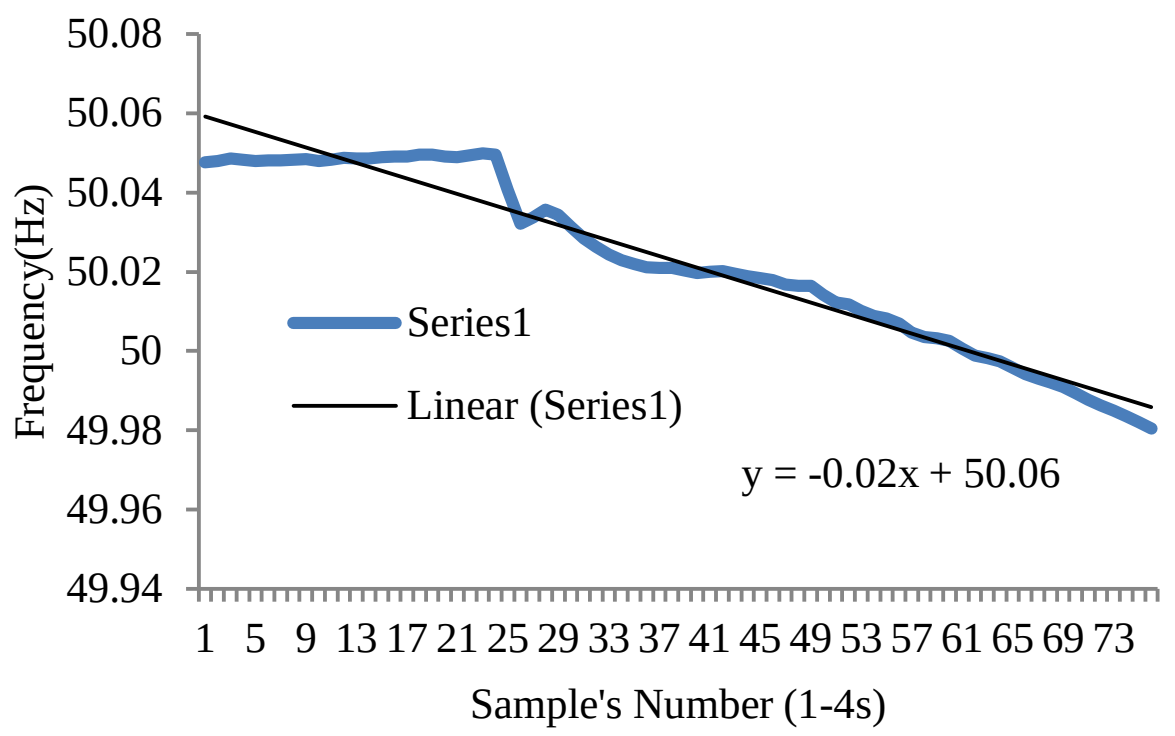

Fig. 8. Frequency curve in the interval from 1 to 4 s

With the given data for the event of a loss of a power plant in the Iranian power system, and considering the power factor is 0.9, the inertia constant of the system is calculated to be 9.66 s:

$$H = \frac{50 \times \frac{170}{19830 \div 0.9}}{2 \times 0.02} = 9.66$$

The actual value of the system inertia constant, in this case was 9.6s. It should be noted that the nominal system frequency is 50 hertz.

## VII. Conclusion

In this paper, an improved method to calculate the inertia constant $H$ of a power system, based on frequency data measured from PMUs, is presented. It is based on the Swing Equation, measured system frequency before and after the disturbance, and the mismatch of real power. The proposed method was applied to data obtained from simulations on the IEEE 39-bus test system and from real-life measurements in the Iranian power grid. The obtained results prove the validity and practicality of the approach.

## VIII. References

[1] A.G. Phadke and J.S. Thorp, "Synchronized Phasor Measurements and Their Applications", Springer, 2008.

[2] S. Chakrabarti, E. Kyriakides, B Tianshu Bi; D. Cai, V. Terzija, "Measurements Get Together," IEEE Power and Energy Magazine, vol.7, no.1, pp.41-49, January-February 2009.

[3] Saeed Khazaee, Mohammadali Hayerikhiyavi, Shahram Montaser Kouhsari, “A Direct-Based Method for Real-Time Transient Stability Assessment of Power Systems”, CRPASE Vol. 06(02), 108-113, June 2020.

[4] M.Shamirzaee†, H.Ayoubzadeh, D.Farokhzad,” An Improved Method for Estimation of Inertia Constant of Power System Based on Polynomial Approximation ’’

[5] G. B. Alinejad, M. Akbari, H. Kazemi, "PMU-Based Distribution Network Load Modelling Using Harmony Search Algorithm”, 17th Conference on Electrical Power Distribution Networks (EPDC), pp 1-6, 2012.

[6] W. H. Karegar,B. Alinejad, “On-Line Transmission Line Zero Sequence Impedance Estimation Using Phasor Measurement Units”, 22nd Australasian Universities Power Engineering Conference (AUPEC) , pp 1-5, 2012.

[7] B. Alinejad, M. Akbari, H. Kazemi, "Detection of Unstable Low Frequency Oscillations Based on PMU Measurements”, Smart Grid Conference (SGC), Tehran, 17-18 Dec, pp 13-19, 2013.

[8] S. Soni,S. Bhil, D. Mehta, S. Wagh, “Linear State Estimation Model Using Phasor Measurement Unit (PMU) Technology”, 9th International Conference onElectrical Engineering, Computing Science and Automatic Control (CCE), pp 1-6, 2012.

[9] P. Wall, F. Gonzalez-Longatt, and V. Terzija, "Demonstration of an Inertia Constant Estimation Method Through Simulation," in Universities Power Engineering Conference (UPEC), 2010 45th International, 2010, pp. 1-6.

[10] S. Mahdavi and A. Dimitrovski, “Integrated Coordination of Voltage Regulators with Distributed Cooperative Inverter Control in Systems with High Penetration of Dg’s,” in2020 IEEE Texas Power and Energy Conference (TPEC), Feb 2020, pp. 1–6.

[11] Peter Wall, Francisco Gonzalez-Longatt, Vladimir Terzija,” Estimation of Generator Inertia Available During a Disturbance”. 2012 IEEE Power and Energy Society General Meeting.

[12] Phadke, J. S. Thorp, and K. J. Karimi, “State Estimation with Phasor Measurements,” IEEE Power Engineering Review, Vol. PER-6, No. 2, pp. 48–48, Feb 1986.

[13] Alinejad, H. Kazemi Karegar,” Online Inertia Constant and Thévenin Equivalent Estimation Using PMU Data”, International Journal of Smart Electrical Engineering, Vol.3, No.3, Summer 2014.

[14] T. Inoue, H. Taniguchi, Y. Ikeguchi and K. Yoshida, “Estimation of Power System Inertia Constant and Capacity of Spinning-Reserve Support Generators Using Measure Frequency Transients,” IEEE Trans. Power Syst., Vol. 12, No. 1, Feb. 1997.

[15] Yuchen Zhang; Yan Xu; Zhao Yang Dong.” Robust Ensemble Data Analytics for Incomplete PMU Measurements-Based Power System Stability Assessment”, IEEE Transactions on Power Systems, 2018.

[16] F. Lingling, M. Zhixin, and Y. Wehbe, “Application of Dynamic State and Parameter Estimation Techniques on Real-World Data,” IEEE Trans. Smart Grid, Vol. 4, No. 2, pp. 1133–1141, Jun. 2013.

[17] Yajun Wang, Horacio Silva-Saravia, Hector Pulgar-Painemal, “Estimating Inertia Distribution to Enhance Power System Dynamics” NAPS Conference 2017, pp1-6.

[18] H. Haggi, W. Sun, and J. Qi, “Multi-Objective PMU Allocation for Resilient Power System Monitoring,” in 2020 IEEE Power & Energy Society General Meeting (PESGM). IEEE, Aug. 2020, pp. 1–5.

[19] Xinan Wang; Di Shi; Jianhui Wang; Zhe Yu; Zhiwei Wang,” Online Identification and Data Recovery for PMU Data Manipulation Attack”, IEEE Transactions on Smart Grid, 2019.